# Temperature manipulation in songbird brain implicates the premotor nucleus HVC in birdsong syntax

Yisi Zhang[1], Jason D. Wittenbach[1,3], Dezhe Z. Jin[1,3*], Alexay Kozhevnikov[1,2,+]

Departments of Physics[1] and Psychology[2] and Center for Neural Engineering[3], Pennsylvania State University, University Park, USA

* Email address dzj2@psu.edu
+ Email address akozhevn@phys.psu.edu

## Significance

Like many animal behaviors, birdsong consists of variable sequences of discrete actions. The ordering of the actions often obeys a set of probabilistic rules. Where and how these rules are encoded in the brain is poorly understood. To address this issue, we locally and reversibly cooled brain areas in songbirds during singing. Mild cooling of the Bengalese finch's brain area HVC – a premotor area homologous to the mammalian premotor cortex – alters the statistics of the bird's song sequences. Our manipulations show that HVC is a critical area for encoding the probabilistic rules of birdsong. These experiments have established a causal link between a brain area and the rules of action sequences through real-time manipulation of intact brain circuitry.

## ABSTRACT

Behavioral sequences of animals are often structured and can be described by probabilistic rules (or "action syntax"). The patterns of vocal elements in birdsong are a prime example. The encoding of such rules in neural circuits is poorly understood. Here we locate the site of song syntax in the Bengalese finch by rapidly and reversibly manipulating the temperature in the song production pathway. Changing the temperature in the premotor nucleus HVC (proper name) alters the transition probabilities between syllables. Most prominently, cooling reduces the number of repetitions of long repeated syllables, while heating increases repetition. In contrast, changing the temperature of the downstream motor area RA (robust nucleus of the acropallium), which is critical for singing, does not affect the song syntax. Computational modeling suggests that temperature can alter the transition probabilities by affecting the efficacy of the synapses in HVC that carry auditory feedback to the motor circuits. The model is supported by a real-time distorted auditory feedback experiment, which shows that perturbing auditory feedback shortens syllable repeats similar to cooling HVC. Taken together, these findings implicate HVC as a key player in determining birdsong syntax.

INTRODUCTION

Many animal behaviors consist of variable sequences of discrete actions. Examples include birdsong (1), whale song (2), and grooming in rodents (3). These behavioral sequences display regularities and structures that are often referred to as "action syntax" (4), in analogy to syntax in language (5). Where and how such syntactic rules are encoded in the brain remains an important unsolved problem.

The songbird is a model for studying the neural mechanisms of vocal sequences (1). Many songbird species sing variable songs (6). One such species is the Bengalese finch, whose song consists of sequences of stereotyped vocal elements called syllables (Fig. 1a,b) (7–9). The ordering of the syllables within the song can change from rendition to rendition. However certain statistical properties of the song, such as repeat distributions (the probability of observing a syllable a certain number of times in a row) and pair-wise transition probabilities (the probability of seeing one syllable following another), are stable over long periods of time (Fig. 1c) (10). Patterns such as these comprise the song syntax.

Singing in songbirds is controlled by a set of linked forebrain nuclei called the song system (11). Among them, the premotor nucleus HVC (proper name) plays a prominent role in defining the fine-scale timing of acoustic features in the song (12–14). It is also the major target of the auditory inputs to the song system (15). HVC projects to the motor area RA (robust nucleus of the acropallium) (11), which transforms the timing signals from HVC and drives learned acoustic features through downstream motor neurons (16).

Where the song syntax is encoded in the song system is not understood. Computational models suggest that HVC contains local circuitry that can implement the probabilistic syllable transitions (17). We test the role of HVC in the song syntax through reversible manipulation of its temperature in singing Bengalese finches. We find that HVC temperature affects both repeat distributions and transition probabilities. In contrast, temperature manipulations in RA have no effect on the song syntax. These observations are consistent with a computational model in which the transition probabilities are determined by integration of the auditory feedback with the intrinsic premotor activity in HVC. The role of the auditory feedback is further supported by a distorted auditory feedback experiment. Taken together, our results show that HVC is a key site for shaping birdsong syntax.

RESULTS

We constructed a miniature thermoelectric Peltier device for reversible manipulation of local brain temperature (13, 18). We cooled or heated HVC and RA to study the effects of temperature on the song timing and the song syntax of the Bengalese finch (Methods, Supplementary Information). Heating consistently showed the opposite effects of cooling. Thus, to simplify the presentation in the rest of the paper, we will discuss the cooling effects and leave the heating data to the figures.

*Effects of cooling HVC on song timing*

Cooling HVC increased the durations of the syllables and the inter-syllable gaps (Fig. 1d,e). The durations of the song syllables were stretched by -2.8±0.9 %/°C (mean ± s.d., percent per degree Celsius, n=34 syllables in 5 birds). Durations of the inter-syllable gaps were stretched by -4.2±2.5 %°C (n=50 inter-syllable gaps in 5 birds). Overall, the amount of stretch in the syllable and gap durations are similar to that seen previously in cooling HVC in the zebra finch, a species whose song consists of fixed sequences of syllables (13, 19).

*Effects of cooling HVC on song syntax*

**Syllable repetition.** In the normal songs of the five Bengalese finches, repeated syllables were classified into two types based on the means and the standard deviations of the repeat number distributions (k-means algorithm, Supplementary Information, Fig. S2) (20). Type I syllables have distributions with means larger than 4 and standard deviations larger than 1, characterizing long and variable syllable repeats; Type II syllables have means smaller than 4 and standard deviations smaller than 1, characterizing a short and nearly fixed number of repeats. In these songs, we identified 8 Type I repeat syllables and 5 Type II repeat syllables.

Cooling HVC consistently shortened the repeats of Type I syllables. An example is shown in Fig. 2, where we show the effects of HVC temperature on a long repeated syllable 'A' of Bird 1 (Fig. 2a,b). This trend is seen in all Type I syllables (Fig. 2c), and the effect is significant on aggregation ($p=1.4\times10^{-10}$, t-test for slopes) and across individual birds (Fig. 2d). In contrast, cooling HVC had minimal effects on the repeat lengths of Type II syllables (p=0.02, t-test for slopes, Fig. 2c). The difference between the two types is significant ($p=6.6\times10^{-8}$, multivariate regression analysis). In the following we will focus on Type I repeat syllables.

A simple explanation for the shortening of syllable repeats by cooling is that the song tempo is slowed while the time course of the repeat bouts – which could be encoded in a different brain area – is unaffected by the cooling. In this scenario, fewer syllables can "fit" in a repeat bout because the syllables and gaps are longer. Under this hypothesis, we can derive an expression for how a repeat distribution should change as a function of temperature for a given amount of syllable/gap stretch (Methods). Using our measured values for syllable/gap stretch, we compared this theoretical prediction to the actual data. One example is shown in Fig. 2e (syllable 'A' from Bird 1 at ΔT=-2°C). The predicted reduction in mean repeat length is significantly smaller than that observed ($p = 1.3\times10^{-11}$, one-tailed t-test). This is true across all Type I syllables (Fig. 2f). Compared to the predicted values, the observed data show significantly larger reductions in repetition as HVC is cooled ($p=2.1\times10^{-5}$, multivariate regression analysis). Therefore, the cooling induced reduction in syllable repetition is not simply a byproduct of slowed song tempo.

**Branch points.** A branch point in the song syntax is a syllable that can be followed by more than one unique syllable in a probabilistic fashion. An example is shown in Fig. 3a, in which syllable 'K' can transition to syllable 'B' with probability 0.68 or to syllable 'D' with probability 0.32 (Bird 3). We identified a total of 12 branch points in the songs of the five birds. The branch points are defined after controlling for syllable misclassification (Supplementary Information and

Methods). In the normal condition, all transition probabilities were stable across days ($p>0.05$, Fisher's exact test).

HVC temperature affects the branch points. In the example shown in Fig. 3a, when HVC is cooled by 4°C, the probability for transition 'KB' is reduced to 0.45, while the probability for 'KD' is enhanced to 0.55. These changes are highly significant ($p = 1.7\times10^{-5}$, Fisher's exact test). The transition probabilities of 8 out of the 12 branch points are significantly dependent on temperature ($p<0.05$, chi-squared test of independence or Fisher's exact test for counts less than 10).

Cooling HVC also affects the variability of the syllable sequences. The randomness of the syllable transitions at a branch point can be quantified by the transition entropy (7, 8, 21) (Methods). A deterministic transition has the lowest entropy, and the value is 0. The most random transition, in which the transitions to all N targets are equally probable, has the maximum entropy, with a value of $\log_2 N$. We calculated the transition entropy for the 12 branch points. Cooling HVC significantly increased the transition entropy in four branch points ($p<0.05$, two-tailed t-test on the slope of the transition entropy versus the temperature), slightly reduced the transition entropy in one ($p<0.05$), and had no significant effects on the rest (Fig. 3b). However, the extent of the cooling effect depends on the transition entropy of the branch points under normal conditions. The branch points with lower transition entropy at baseline temperature tend to increase their transition entropy more when HVC is cooled (Fig. 3c). At the same time, all of the branch points that showed either no change or a slight increase in entropy all had baseline values that were close to the maximum possible value to begin with. In other words, cooling HVC increases the randomness of the syllable transitions that are close to being stereotypical in normal conditions, but does not strongly affect the transitions that are already random. Overall, this shows that cooling HVC increases the randomness of syllable transitions.

Besides the changes of the transition probabilities, we also observed appearance of two novel transitions in one bird (Bird 7) at $\Delta T=-4°C$, the lowest cooling achieved in our experiments (Supplementary Information). In one case, the stereotypical transition from syllable 'H' to syllable 'G' in the normal condition became variable, and a new transition from syllable 'H' to syllable 'B' appeared with a probability 0.31. In the other case, the normal transitions consisted of syllable 'B' transitioning to syllable 'D' with a probability 0.94 and to syllable 'C' with a probability 0.06. At $\Delta T=-4°C$, the probability for 'BD' was reduced to 0.69 and the probability for 'BC' increased to 0.21. An additional transition 'BF' appeared with a probability 0.1. The appearances of these novel transitions can be seen as an additional way of randomizing the syllable sequences as HVC is cooled. However these events were rare and only observed in the lowest HVC temperature achieved in the experiments.

*Effects of cooling RA*

To assess the importance of HVC for the song syntax relative to other areas in the song system, we directly manipulated the temperature of RA. RA is a major motor area for birdsong production, and is directly innervated by HVC (11). Unlike HVC, RA is located deep (2 mm) below the brain surface, and its temperature is harder to manipulate. We succeeded in cooling RA in two birds, using thermally conductive probes (18).

Changing RA temperature had only mild effect on song timing and syntax (Fig. 4). Cooling RA stretched the syllables by -0.9±0.1%/°C (n = 22) and the gaps by -0.5±0.3%/°C (n=34) (Fig. 4a). The sizes of these changes are about one third of what we observed when HVC was cooled. Moreover, cooling RA led to cooling in HVC, with the change of temperature about 32% of that in RA (Supplementary Information, Fig. S3). These observations imply that the effects of RA temperature on the syllable durations could be explained by the collateral temperature change in HVC, which is verified statistically by comparing the fractional stretch with cooling RA and the stretch with cooling HVC multiplied by a factor of 32% (t-test, p = 0.94). However, the stretch of gaps with cooling RA yields slightly less effects than that can be accounted for by the collateral temperature changes (t-test, p = 0.01).

In the songs of the two RA cooled birds, there were four Type I repeat syllables (Supplementary Information, Fig. S2) and seven branch points. The mean repeat lengths of these syllables did not change significantly when RA was cooled (Fig. 4c, p>0.05, t-test on slopes). The sizes of the changes were also significantly smaller than those observed in HVC cooling (Fig. 4d, $p=2.2\times10^{-7}$, multivariate regression analysis). Among the seven branch points, two had transition probabilities that were affected by temperature (p<0.05, Fisher's test). However, no significant trends in the changes of the transition probabilities were observed as RA was cooled, regardless of the transition entropies in normal condition. (Fig. 4e, p>0.05, t-test). Thus, the RA temperature manipulation has minimal impact on the song syntax.

*Modeling the effects of HVC cooling on the song syntax*

Our experiments show that cooling HVC impacts the song syntax far more than cooling RA. One interpretation is that the song syntax is encoded within HVC. To show this is plausible, we build a biologically detailed computational model and demonstrate that manipulating HVC temperature changes the transition probabilities between the syllables as observed in the experiments.

The model is modified from a previous model of generating variable birdsong syntax (17). In the model, syllables are encoded in chain networks of RA projecting HVC ($HVC_{RA}$) neurons. Spike propagation through a chain drives RA and downstream motor neurons to produce an associated syllable, as shown in Fig. 5a (12, 14, 16, 17, 22). Probabilistic transitions between syllables are encoded in the branching connections between their respective chains. The $HVC_{RA}$ neurons are subject to a global feedback inhibition mediated by the inhibitory HVC interneurons, which enforces a winner-take-all competition at the branch points and ensures that only one chain is active after a brief transition period (17).

Auditory feedback is known to affect Bengalese finch song syntax in real-time (21) To account for this, we extended the model to include syllable-specific auditory feedback that influences the HVC dynamics via synapses that innervate the chains of $HVC_{RA}$ neurons (20, 23) . The auditory inputs for a given syllable have different strengths on different chains, and thus provide additional syllable-specific excitation that biases the transition probabilities. Finally, in order to reproduce repeat distributions with peaks located at a large number of repetitions (i.e. the Type I syllables), we add short-term synaptic depression to the synapses carrying the auditory feedback.

This depression implements stimulus-specific adaptation, which is sufficient for describing the statistics of long-repeating syllables and has been observed in syllable-specific auditory responses within HVC (20).

To assess how cooling affects the model dynamics, we introduced temperature dependence of the efficacies of neural and synaptic mechanisms in the model, using typical values found in the literature for various neuron types (Methods) (24–31) When cooled, the neurons become more excitable (24), the synaptic transmissions are reduced (30, 31), and synaptic depression is stronger (25, 31) . We also assumed that cooling affects the auditory synapses more strongly than the synapses between HVC neurons.

The model reproduces the cooling effects on the song tempo and the song syntax. Similar to the experimental results, cooling stretches syllables in the model (Fig. 5b). The amount of duration stretch is -3.8 %/°C.

Cooling also affects the repeat dynamics of long-repeating syllables. In our model, such syllables are created when auditory feedback from a syllable targets the chain that encodes this syllable, biasing the network toward repetition. Due to the reduction of the drive from the auditory feedback, as well as a faster depression on this drive, there are fewer repetitions of such syllables as HVC is cooled (Fig. 5d,e). This effect mirrors the similar trend seen in the experimental data for Type I syllable repeats. As with the experimental results, the reduction of repetition due to cooling (Fig. 5d, red line) is much more than expected from an alternative model where repetition is reduced solely due to syllable stretch (Fig. 5d, purple line).

The idea that auditory feedback can bias transition probabilities can be extended to the branch points, and can be used to explain the increase of transition entropy when HVC is cooled. For this we assumed that the branching chain networks produce nearly equal transition probabilities at the branch points due to similar connection strengths to the branching chains. In this default state of the intrinsic HVC network, the transition entropy is near the maximum. Through preferential targeting of the auditory feedback to a subset of the branches, the transition probability can be biased to that subset of transitions, which lowers the transition entropy. This is illustrated with a network consisting of three chains, representing transition from syllable 'A to syllable 'B' or to syllable 'C' (Fig. 5f). Since cooling reduces the efficacy of the auditory synapses more than the synapses between HVC neurons, the network dynamics reverts to the default state with nearly maximum transition entropy. This effect of cooling on transition entropy is shown in Fig. 5g, in which we show the results for the three cases of integrating auditory feedback to the chain network shown in Fig. 5f. In the first case, none of the chains receive auditory feedback. This intrinsic network produces transition probabilities 0.57 for 'AB' and 0.43 for 'AC'. When cooled, the transition probabilities fluctuate in a random way, but the transition entropy remains near the maximum (black curve). In the second case, the auditory feedback from syllable 'A' is applied to chain-B, increasing the 'AB' transition probability to 0.89 in the normal condition. When cooled, the reduction in the auditory feedback increases the transition entropy (blue curve). In the third case, the auditory feedback is applied to chain-C. Similar results were obtained (purple curve).

*Effects of distorted auditory feedback on syllable repetition*

Previous experiments have shown that distorted auditory feedback influences the syllable transition probabilities in the Bengalese finch (21). To see whether the auditory feedback also plays a role in syllable repetition, as suggested by our model, we performed a real-time distorted auditory feedback (DAF) experiment in normal conditions on one of the birds used in the cooling experiments. We targeted a Type I repeat syllable (Fig. 6a). A brief white noise (~50ms) was presented during each ongoing repeat of the syllable (Fig. 6a). Only repetitions where the song was not terminated after the white noise were used in the analysis to avoid conflating song interruption with a change in repetition syntax. Similar to the effect of cooling HVC, the length of the syllable repeats is significantly reduced by DAF (Fig. 6b, $p=5.1\times10^{-13}$, t-test). The distribution of repeat length with DAF is not different from that of the same syllable with cooling by 2°C (Fig. 6c, $p=0.07$, Kolmogorov-Smirnov test). This observation supports the role of auditory feedback in producing syllable repetition.

DISCUSSION

Sequences of discrete actions are prevalent in animal and human behaviors (1–4). Speech and language in humans are perhaps the most elaborate form of action sequences. The syntax of human language is characterized by recursive structures (5), which is lacking in action sequences in animals (32). Nonetheless, action sequences in animals have regularities that can be concisely summarized in a set of probabilistic syntactical rules, or action syntax (2–4, 6, 9, 32).

Where and how syntactic rules are encoded in the brain remains a critical unsolved problem in systems neuroscience. In our work, we have established causal links between areas in the song system and the song syntax of the Bengalese finch using a rapid, reversible cooling technique on intact brains. This is an important advance compared to other techniques. In animals, lesioning has been the main method for locating sites controlling action syntax (3, 33). However, by causing irreversible brain damage, lesions may lead to neural compensation, which obscures the true role of the lesioned site in the intact brain. In the human brain, the loci controlling syntax have been searched for using functional magnetic resonance imaging (34), positron emission tomography (35), and brain lesions in aphasic patients (36, 37). Imaging techniques, while non-invasive, are correlational and lack direct causal proof that a brain site is enforcing syntactical rules.

It is often thought that motor sequence control is organized hierarchically (38). Applied to birdsong, one might image that different aspects of the song, such as sound features, syllables, and song syntax are encoded in different areas in the song system that are linked in a feed-forward pathway. Indeed, single unit recordings (12, 39) and cooling experiments in zebra finches (13) have shown that that HVC encodes the temporal dynamics at the syllable level, while RA encodes moment-to-moment sub-syllabic features (16). Given that the HVC to RA projection is the major feed-forward pathway for song production (11), hierarchical control does apply at the level of syllables and sub-syllable features. A natural extension of this idea is to assume that the sequencing of the syllables is determined in an area presynaptic to HVC, such as the thalamic nucleus Uva (nucleus uvaeformis) and the forebrain nucleus NIf (the nucleus interface of the nidopallium) (11). These areas have been implicated in syllable sequencing in lesion studies. Lesioning Uva in the zebra finch disrupted song sequence stereotypy (40, 41).

Lesioning NIf, which provides auditory inputs to HVC (15, 42), affected the syllable sequencing in the Bengalese finch (33). Thus, Uva and NIf appear to be the prime candidates for the site of encoding song syntax. However, our cooling experiments show that HVC is directly involved in controlling the song syntax and the notion that HVC merely follows sequencing commands from upstream areas is incorrect. Our computational model further suggests that HVC may be the main site for encoding song syntax. The proposed mechanism is one where feedback signals from NIf or Uva are integrated with the intrinsic HVC network for generating probabilistic syllable transitions. In this view, NIf and Uva play a secondary role in the song syntax by providing feedback inputs to HVC.

Cooling HVC affects inputs to downstream nuclei in the song system, including RA and Area X, the basal ganglia circuit in the song system (43). A natural question is whether these input changes play a role in the syntax modification. We directly showed that cooling RA, which affects RA's intrinsic dynamics as well as the inputs to RA from other areas, has minimal effects on the song syntax, ruling out the involvement of RA and its inputs in determining the song syntax. A previous experiment showed that removing LMAN (the lateral magnocellular nucleus of the anterior nidopallium), the output station of Area X that links Area X to RA, had no impact on the song syntax in the Bengalese finch (44). This suggests that Area X and LMAN play no role in determining the song syntax. Therefore, changing inputs from HVC to area X should have no effects on the song syntax. Taken together, these results suggest that song syntax is determined within HVC.

Our computational model provides a plausible explanation for the effects of HVC cooling on the song syntax. In the model, the allowed syllable transitions are encoded in the branching connections between the syllable-encoding chain networks of $HVC_{RA}$ neurons. This intrinsic HVC network produces nearly equal transition probabilities at the branch points. To make the transition probabilities biased toward one branch, auditory feedback preferentially targets that branch and enhances the transition probability by providing additional excitatory inputs. A special case of this mechanism is production of long repeats of a syllable. Auditory feedback to the repeated syllable is initially strong, making the repeat probability close to 1. As the syllable repeats, the auditory synapses are weakened due to synaptic adaptation, which leads to eventual termination of the repetition. This mechanism explains statistical properties of syllable repeats in Bengalese finch songs as well as the reduction of syllable repeats by deafening (20). It is also consistent with our observation that altered auditory feedback during syllable repeats reduces repetition. Cooling HVC reduces the efficacy of the auditory feedback, leading to reduction of syllable repeats. At branch points, cooling reduces the bias in transition probabilities due to the auditory feedback, and hence enhances the transition entropy. This result agrees with a previous observation that altering auditory feedback increases transition entropy in the songs of the Bengalese finch (21). Thus our cooling results on the Bengalese finch are consistent our HVC-based model. The interaction of the inputs to HVC and the intrinsic HVC dynamics in our model could be the mechanism for the observed effects of Uva lesion (40, 41) and NIf lesion (33) on the song syntax.

It should be noted that, in our model, cooling has two important effects on how auditory feedback biases transition probabilities: (1) the auditory feedback drive is weakened relative to the chain-to-chain drive; (2) synaptic depression is stronger, leading to greater stimulus-specific adaptation. The result of both of these effects is reduced efficacy of the auditory feedback to bias transition probabilities. Without auditory feedback, we only see small random changes in syntax, most likely due to changes in neural activity patterns caused by slower propagation of the signal within the chains. Thus the two effects stated above are the driving factors behind the significant changes in syntax that we observe in the model. Both of these effects can independently lead to a reduction of syllable repeats. However, at branch points, only the former can cause increases in transition entropy (data not shown). This is due to the fact that the stimulus-specific adaptation only has an opportunity to build up to significant levels with the repeated activation of the same auditory feedback signal, as seen during repetition but not at branch points.

In our experiments, cooling affected Type I but not Type II syllable repeats. Type I repeats are characterized by long and variable repeat numbers from rendition to rendition, while Type II repeats have short and nearly fixed repeat numbers. These distinctions can be explained in our model by the relative importance of the auditory feedback compared to the HVC intrinsic connections (20). When the auditory feedback is weak, syllable repetition relies on the self-connection of the chain network for the repeat syllable. The repeat probability is nearly fixed, and the value is smaller than one. The distribution of repeat numbers is a decreasing function, and the most probable repeat number is one. The mean and variance of repeat numbers is thus small. In contrast, when the auditory feedback dominates, the repeat probability is nearly 1 until the auditory feedback is weakened by synaptic adaptation. This enables a Gaussian-like repeat number distribution, with the most probable repeat number much larger than one. The mean and the variance of such a distribution are large. When deafened, Type I repeat syllables are much more affected than Type II repeat syllables (20). This mirrors the effects of HVC cooling, which reduces the effectiveness of the auditory feedback. Another way of generating a short and fixed number of repeats is using a "many-to-one mapping" from chain networks to syllables (9, 17). In this mechanism, several HVC chains are connected in a row, and all of them encode the same syllable by driving the downstream motor areas in the same way. Because this mechanism relies on HVC's intrinsic dynamics, cooling and deafening would not have much effect on such repeat dynamics.

Cooling HVC affected the transition probabilities at the branch points. Cooling randomized the transitions that were more stereotypical in the normal condition. Random transitions remained random. Overall, cooling HVC makes the syllable sequences more variable. In our experiments, HVC cooling left the rules of syllable transitions (i.e. the allowed and forbidden syllable transitions) largely intact. However, novel transitions did occur at the lowest cooling condition, demonstrating that the rules of the song syntax can also be modified by HVC cooling. This suggests that the rules and transition probabilities of birdsong syntax are likely encoded together within HVC. In our model, the rules of the transitions are encoded in the branching patterns of the syllable-encoding chains. The transition probabilities are determined by the winner-take-all dynamics at these branch points. These connection patterns could be hard to perturb by HVC cooling, which might explain why the transition rules were rarely perturbed in our experiments, unlike the transition probabilities. Extreme cooling of HVC weakens the connections between the HVC neurons, which could give an opportunity for the transitions deviate from the branching connections and create novel transitions.

When changing HVC temperature, the surrounding areas are also inevitably affected. To estimate the effects of this collateral cooling, we measured the temperature gradient in the HVC vicinity (Supplementary Information). The primary forebrain auditory area, field L, is located approximately 1.0-1.5 mm ventral to the center of HVC (45). When HVC temperature was varied between -4.3°C and 2°C, the temperature at this distance changed between -2.5°C to 1.5° (Supplementary Information). The temperature in NIf, which is about 2.5-3.0 mm ventral to the center of HVC (45), was changed between -1.7°C to 0.5°C. It is conceivable that such temperature changes in this auditory area might impact the song syntax. However, we found that RA cooling affected NIf temperature similarly to HVC cooling (Supplementary Information, Fig. S4c), and in this case, we observed very little effect on the song syntax. This suggests that the collateral temperature change in NIf was not important in our experiments. Song interruption due to the distorted auditory feedback was similar in the normal condition and when HVC was cooled by 4°C, suggesting that the amount of collateral cooling in Field L did not impair hearing. Future experiments are needed to directly probe whether the amount of temperature change in Field L seen in our experiments has any impact on the song syntax. Uva is a deep thalamic nucleus (3.0-3.5 mm ventral to the center of HVC) (45), and is closer to RA than to HVC. Therefore, it is unlikely that Uva was affected by HVC cooling and responsible for the syntax changes in our experiments.

When HVC temperature is changed, inter-syllable gaps stretch more than song syllables for Bengalese finches (p=0.016, multilevel model test), and the difference between the stretch of the syllables and gaps is larger than the difference reported for zebra finch (13, 19). This difference might be related to the differences between the syntax of the zebra finch song and the Bengalese finch song and their underlying neural mechanisms. The gaps in a Bengalese finch song are more variable in duration compared to the syllables (Supplementary Information, Fig. S6). In our model, the mutual competition among the candidate branching chains occurs during the gaps. This competition process could cause greater variability in the duration of gaps and may be more affected by the temperature change, resulting in a greater stretch in the gaps compared to song syllables.

The finding that reversible manipulations of HVC circuitry alter the song syntax may be consistent with several alternative syntax mechanisms, including neural feedback-mediated syntax (46) and co-articulation phenomenon (47) . Given the complexity of the neural circuitry and the level of current understanding of biological circuits, further experiments will be needed to decisively establish the neural mechanism that is consistent with all experimental evidence. Additional questions about birdsong syntax remain, such as the interaction between the motor circuits in the two hemispheres. It has been shown in zebra finch that the coordination of the two hemispheres is important for controlling the song tempo (13, 48); however, the role of the two hemispheres in control remains unknown.

To characterize the syntax changes in our experiments, we analyzed the changes in the syllable transition probabilities. Although the changes in transition probabilities unambiguously indicate the syntax change, transition probabilities alone do not fully capture the statistical properties of the syllable sequences in the Bengalese finch (9, 49, 50). It will be interesting to see whether more subtle changes in the syntax can be detected with statistically correct representations of the song syntax, such as the partially observable Markov model with adaptation (9) and the long-range order of the song sequence (51).

In summary, our experimental and computational findings demonstrate the close inter-dependence of the neural mechanisms behind both syntax and timing and suggest that brain area HVC is the key site for birdsong syntax generation. This result restricts the plausible birdsong syntax models to the models where the tempo and the syntax mechanisms are inter-dependent, and HVC is strongly involved in both. This finding will motivate further theoretical and experimental studies elucidating the neural mechanisms of temporal order.

METHODS

*Manipulations of HVC temperature*

We constructed a device for reversible, local brain temperature manipulation that is similar to a device recently described (18). Briefly, the device consists of a Peltier thermoelectric element, gold-plated contact pads and a heat sink. Because HVC is close to the skull, the contact pads were placed on top of the dura. The device was attached to the skull using dental acrylic. Application of direct current to the device changed the brain temperature in the vicinity of the pads. Temperature of HVC was monitored with a miniature thermocouple inserted in the vicinity of HVC at the depth of 0.5mm (Supplementary Information). All procedures are carried out in accordance with the protocol approved by the local Institutional Animal Care and Use Committee.

*Manipulations of RA temperature*

The design of the RA cooling probes was similar to Long's 2008 paper (13). The cooling device was implanted in an anaesthetized bird with probes bilaterally inserted into RA. In order to measure RA temperature, a thermocouple was inserted anterior to the probe from a different angle to reach RA. Another thermocouple was inserted in the ipsilateral HVC at a depth of 0.5mm underneath the brain surface. Both RA and HVC temperatures were recorded simultaneously for different applied currents.

To record changes in spontaneous RA activity due to temperature, a carbon fiber electrode (Carbostar-1, Kation Scientific) was inserted into RA in an anaesthetized bird. The neural signal was amplified by 100 times and filtered with a cut-off frequency at 2kHz (FLA-01, Cygnus Technology) and then fed into an A/D converter (USB 6211, National Instrument) with which the signal was digitized and stored.

*Distorted auditory feedback*

A targeted syllable was chosen, and the first 15-20ms of that syllable were used as the template in the algorithm for on-line syllable detection. The real-time auditory feedback system is described in Skocik et al (52). Once the syllable was detected, the system generated a white noise pulse of 30-60ms long, and 70-80dB sound pressure level. Around 30%-50% of song bouts were randomly selected to receive DAF. For the selected bouts, DAF was applied to all instances of the repeat syllable targeted.

*Data acquisition*

Undirected songs were recorded (16-bit resolution, sample rate 40 kHz) and analyzed off-line. Segmentation of the song into song syllables and inter-syllable gaps is done based on the envelope amplitude crossing a threshold. Syllable identification is first done automatically using custom sorting algorithm (9). The results of automatic sorting are manually verified by visual inspection of the syllable spectrograms. To eliminate the possible long-term changes in the song, data comparison is done between the control data and the test condition data collected on the same day. Time intervals when brain temperature is altered are interleaved with the time intervals when the brain temperature is not changed to ensure that temperature manipulation does not cause irreversible song changes.

*Fractional stretch of the duration of song elements*

To assess the effects of temperature changes, we compared the durations of the song syllables, the durations of the inter-syllable gaps and the sequences of song syllables in the songs under control condition and when the HVC or RA temperature was altered. At each temperature, the mean duration of each song element (each type of song syllable and each type of inter-syllable gap) was computed. For each song element, a least squares linear fit of the mean duration vs. temperature drop was performed; data at each temperature were weighted by the inverse square of the standard deviation. The slope of the least-squares fit was divided by the mean duration of the song element in the control condition to give a fractional stretch of the song element, expressed in per cent per Celsius (%/°C).

*Analysis of temperature effects on the number of syllable repeats*

We computed the mean lengths of the repeats for all the repeating syllables at different temperatures. To separate the long and variable repeats from the short and fixed-numbered repeats (or Type I and II as defined in the Results), we took the mean and standard deviation of the repeat length in the $\Delta T=0$ group, and used the z-scores of them to perform k-means clustering. Only repeats from the Type I cluster were used in the analysis of the temperature effects. To verify that the syllable repeats we investigated for the RA cooling analysis are also Type I repeats, we predicted the classification of these repeats based on the K nearest neighbor classifier constructed from the repeats collected in the HVC cooling experiment (53). The z-scores of the measures (mean repeat length and standard deviation) were calculated using the same center and normalization coefficient. The classifier construction was performed using the 'fitcknn' function and the prediction was performed using the 'predict' function in MATLAB. Some long syllable repeats had a bimodal distribution with a sharp peak at 1-2 number of repeats and a broad peak with a long tail. These are mixtures of Type I and Type II repeats (20). For these syllables, we discarded the repeat lengths that are equal or smaller than 2 and used the Type I parts of the distributions.

To test if the shortening of the repeat length when HVC was cooled is simply due to the stretching of syllable duration without change in the time spans of the repeat bouts, we used the following analysis. In this hypothesis, we have $D(\Delta T) = D(0)$, where $D(\Delta T)$ is a random variable representing the total duration of a repeat bout. We also assume that the stretch in the syllable duration is linear in the temperature, as suggested by the data: $L(\Delta T) = L(0)(1 - \alpha \Delta T)$. Here, $\alpha$ is the fractional stretch per unit change in temperature; we assume that the stretch of an element (syllable+gap) is approximately the average of the syllable and gap stretch, which is about $\alpha = -3.5\%/°C$. Fluctuations in L are much smaller than those in D, so we assume that syllable duration is a non-random quantity. The resulting distribution of the repeat length at any $\Delta T$ can be calculated as $N(\Delta T) = \frac{D(\Delta T)}{L(\Delta T)} = \frac{D(0)}{L(0)(1-\alpha\Delta T)} = \frac{N(0)}{1-\alpha\Delta T}$. Therefore, the repeat length distribution at $\Delta T$ can be generated directly from that at $\Delta T = 0$ by rescaling by a factor of $\frac{1}{1-\alpha\Delta T}$.

*Analysis of temperature effects on the branch points*

To construct the transition probability matrix, we counted the number of transitions from syllable i to j (i≠j) and divide it by the total number of transitions from syllable i, so that the elements in the transition probability matrix of each row sum up to 1. The end of song was identified if the silence following a syllable was longer than 1s or if an introductory note was followed. A start/end state was introduced so that the sequences were cyclic. The introductory notes were merged into the start/end state to exclude them from the sequence analysis. To separate the effect on repeats from the effect on branch points, we did not count self-repeating transitions and treated the repeats as a single "repeated syllable". Thus, the diagonal elements of the transition matrix were set to zero. We did not include the transitions to the start/end state in the analysis of the branch points and the transition entropy. We used a fairly conservative approach to estimate the level of noise in transition counts due to misclassification of syllables (see Supplementary Information for details). The branch points were identified after we correct for the estimated misclassification rate, so that only transitions with a probability above the noise level are considered. With this criterion, only two cases of novel transitions were observed at the extreme condition ΔT=-4°C. Novel transitions were not taken into account in our analysis of the branch points.

*Analysis of sequence variability*

We approximate the song sequence as a Markov chain in which the selection of a syllable is only dependent on the immediately preceding one. We used transition entropy, defined as $H_i = -\sum_j P_{ij} \log_2 P_{ij}$, to quantify the variability of the transitions at the i[th] branch point. We excluded the novel transitions in the analysis of transition entropy. To separate the entropy change due to the change of repeat length, we excluded repetitions in this analysis as well. The transition entropy was further normalized by the maximum value of entropy possible for the corresponding number of transitions, $N$: $\log_2 N$.

*Analysis of DAF effects*

The number of syllable repeats with distorted playback was counted by dividing the total span of the repeat bout by the average duration of the syllable and gap combined. This average was computed using the unperturbed songs. Repeat bouts interrupted by DAF were excluded.

*Statistical analysis*

To assess whether the repeat length significantly depends on HVC or RA temperature, we performed two-tailed t-test on the slopes of the lines fitted to the temperature and the repeat length. In this analysis, the repeat length was centered by subtracting the mean repeat length of the syllable in all temperature conditions. Type I and Type II repeat syllables were analyzed separately.

To see whether the reduction of repeat length with HVC cooling is more significant for Type I than for Type II repeat syllables, we carried out multivariate regression analysis (54). In the regression analysis, the centered repeat length $R_c$ is expressed according to the equation $R_c = \beta_0 + \beta_1 \Delta T + \beta_2 C + \beta_3 C \Delta T$, where $\Delta T$ is the change in HVC temperature, and $C$ is set to 0 for Type I syllable data and 1 for Type II syllable data. $\beta_0, \beta_1, \beta_2, \beta_3$ are the regression coefficients. The p-value was computed for rejecting the null hypothesis $\beta_3 = 0$, which would suggest that the slopes of the regression lines for $R_c$ and $\Delta T$ are the same for Type I and Type II repeats. This analysis was carried out using the multiple linear regression function (lm) in the statistical package R.

We used t-test to see whether the means of two distributions of the repeat lengths at a given HVC temperature are significantly different. One distribution is from the actual data, and the other is generated from a model in which the reduction in repetition is due to the stretches of the syllables and gaps. We used the multivariate regression analysis, similarly as described above, to test whether the reduction of repetition with HVC cooling is significantly different in the actual data than in the model generated data.

The significance of changes in transition probabilities were tested with chi-square test for independence in tables, or Fisher's exact test if some counts of transitions were smaller than 10 (55). For the transitions from one syllable, a contingency table was constructed by counting the number of transitions to all target syllables. Each row of the table represents one target syllable, and each column represent different conditions. For testing the temperature effects, each column contained the counts to the target syllables at one temperature. For testing the day-to-day fluctuations, each column contained the counts observed in one day. Such tables were used to test the null hypothesis that the transition probabilities remain the same in different conditions.

The significance that the transition entropy changed with the HVC or RA temperature was assessed using two-tailed t-test on the null hypothesis that the slope of the line fitted through the data is zero. To assess the dependence of these slopes on the transition entropies at $\Delta T=0$, a line was fitted through data consisting of these slopes and the corresponding transition entropies at $\Delta T = 0$. Two-tailed t-test was used to test whether the slope of the fitted line is significantly non-zero.

To test if changes in the durations of the syllables and gaps with the temperature in RA can be explained by the collateral effect in HVC, we tested the null hypothesis that the fractional stretch of the syllables (or gaps) by per °C temperature change in RA is equal to 32% of that in HVC. The percentage per °C change in duration was obtained by calculating the slope of the fractional stretch with respect to temperature. The slope with cooling in HVC was then multiplied by a factor of 0.32. The significance of the difference in means in the two groups of slopes was computed with the t-test for unequal sample sizes and unequal variances. Multivariate regression analysis, as described previously, was used to show that reduction in Type I repetitions was stronger for HVC cooling than for RA cooling.

To assess the DAF effect on the repeat length, we carried out t-test on significance of the differences in the mean repeat lengths between the control group (without DAF) and the DAF group. We used Kolmogorov-Smirnov test to assess the significance of the difference in the repeat length distributions with HVC cooling and with DAF.

To test whether there are significant differences in stretches of syllables and gaps as HVC is cooled, we carried out multilevel model analysis in which bird identities were taken into account. The fit mixed-effect model (fitlme) in MATLAB was used for the analysis. In this model, the response variable is the fractional stretch, while the predictor is whether it is a gap (1) or a syllable (0) and the bird identities (#1,2,3,4,7) are treated as random effects. The p-value of the slope reflects whether the null hypothesis that the stretches on syllables and gaps are the same can be rejected.

The criterion for significance in all tests was set at $\alpha = 0.05$.

*Modeling the effects of temperature*

**Original model for HVC with adapting auditory feedback.** The original model is described elsewhere (20), but we give a brief outline here. The model is a detailed biophysical model of the neural circuit within HVC based on the branching synfire chain model. A chain of RA-projecting HVC neurons encodes each syllable. Each chain is made up of sequentially ordered groups of these neurons where each group makes all-to-all excitatory synaptic connections to the neurons in the next group. This connectivity promotes sequential activation of the groups within the chain that drives downstream motor production of the associated syllable. Connections from the last group in a given chain "branch" to make all-to-all excitatory connection to the first group of any other chain that encodes a syllable that can follow the syllable of the previous chain. Self-connections are possible and allow for repetition. Global inhibition is implemented by a group of local inhibitory interneurons and supports lateral inhibition between the chains of excitatory neurons. Thus when activity reaches the end of one chain, a winner-take-all competition ensues between the subsequent chains in the branching pattern, so that only one syllable is chosen to follow the previous syllable. Syllable transition probabilities are thus determined by the relative strengths of the branching chain-to-chain synapses. In this way, HVC encodes both the syllable tempo/identity (via the chains) as well as the transition probabilities (via the branching patterns of connectivity between the chains). When simulating the model, a small amount of noise is added in the form of Poisson spike trains incident on each neuron with randomly generated spike times and strengths. This allows for different chains to win the competition at the branch points of different iterations of the same simulation. All neurons are conductance-based models with currents chosen to reproduce salient features of these populations seen in experiments. Synapses implement instantaneous pulse-coupling.

Additionally, syllable-specific auditory feedback (from a nucleus such as NIf) can provide targeted excitatory drive to individual chains. This feedback is modeled as a train of excitatory spikes with Poisson statistics incident of the neurons within a chain. When a given syllable is being produced, all other chains can receive such a drive, starting after a motor-to-sensory feedback delay. Each syllable-to-chain drive can have a different strength set by different synaptic weights. When we say that "Syllable X provides auditory feedback to chain Y", this implies that when syllable X is being produced, only chain Y receives such an external drive with non-zero strength. This provides auditory feedback control, as this strength of this extra drive will bias chain-to-chain transition probabilities. Finally, stimulus-specific adaptation of this sensory feedback signal is modeled as short-term synaptic depression of the synapses carrying the external feedback drive to HVC. The choice of this mechanism in the original model was simply one of convenience – there is no experimental evidence that this is the correct mechanism or even the proper location for source of such adaptation. However it serves the purpose of a phenomenological model that reproduces the properties of stimulus-specific adaptation. For a full description of the model, see (20).

**Temperature dependence.** The model neurons used in for these simulations are single- or two-compartment conductance-based models. Across neuron types, cooling increases the membrane time constants, broadens spikes, and increases the intracellular resistance. In the literature, the temperature dependence of many parameters is quantified by a $Q_{10}$ value – the fractional change in the quantity when the temperature is increased by 10 °C. This gives rise to the relationship: $R(T) = R(T_0)Q_{10}^{(T-T_0)/10}$, where $R(T)$ is the value of the parameter at an arbitrary temperature, $T$, and $T_0 = 40$ °C is the baseline temperature at which the parameter values are known. There are no detailed measurements of HVC neuron properties at different temperatures. We therefore modeled the temperature dependence for our model neurons by choosing $Q_{10}$ values that are representative of those found in the literature, which have been measured for retinal, cortical, and hippocampal neurons. Maximum conductance: $Q_{10} = 1.5$ (24, 26); open/close rates for ion channels $Q_{10} = 2.0$ (26); intracellular resistance $Q_{10} = 0.75$ (28); intracellular calcium concentration decay time constant $Q_{10} = 0.7$ (27); synaptic current time constant $Q_{10} = 0.75$ (29, 30); synaptic maximum conductance – locally originating synapses: $Q_{10} = 1.4$ (29–31), auditory feedback synapses: $Q_{10} = 3.2$; synaptic depression recovery time constant $Q_{10} = 0.45$ (30, 31). The reversal potentials of ionic currents are proportional to the absolute temperature (measured in Kelvin) according to the Nernst equation. Since it must be a fraction between 0 and 1, the synaptic release probability (which also plays a role in the depression dynamics) cannot have its temperature dependence modeled by a $Q_{10}$ value. Instead, we used a simple sigmoid to keep it in the appropriate range: $\alpha(T) = \left[1 + \frac{1-\alpha(T_0)}{\alpha(T_0)} Q^{T-T_0}\right]^{-1}$, where $\alpha$ is the depression strength and $Q$ is a parameter similar to the $Q_{10}$ value that determines the strength of the temperature dependence. We used $Q = 1.25$.

**Network configuration.** To examine how temperature changes affect the syllable duration as well as the repetition statistics in this model, we used a simple network. This network consisted of three chains: call them chains A, B, and C. The end of chain A is connected to the beginning of chain B. The end of chain B is connected back to the beginning of chain B as well as to the beginning of chain C. Activity is always initiated at the beginning of chain A by injecting external current into the neurons in the first group of the chain. This activity then flows down chain A and into chain B. When the activity reaches the end of chain B, a winner-take-all competition between the beginnings of chain B and chain C ensues and activity continues in only one of the chains. In this way, activity repeats in chain B some number of times before continuing to chain C, generating sequences of the form $AB^nC$ (where $B^n$ denotes n copies of B). This represents one bout of repetition of syllable B, preceded by A and followed by C. Auditory feedback from syllable B is targeted back to chain B. This feedback initially sustains repetition of B. As stimulus-specific adaptation sets in, the auditory feedback weakens and continued repetition becomes less likely. This implements long-repeating syllables with a most probable number of repetitions greater than 1. Without the adaptation, the dynamics are Markovian and the most likely number of repeats is always 1. This simulation is run multiple times at multiple temperatures. Syllable stretch is analyzed by computing the average amount of time it takes the neural activity to travel from the middle of chain C to the end of chain C (to avoid any effects of auditory feedback). Repeat number distributions are computed by the fraction of simulations on which chain B repeated a given number of times before neural activity moved on to chain C.

To analyze branch point dynamics, a slightly different network is used. There are three chains, again called A, B, and C for convenience. The end of chain A now makes branching connections to the beginnings of chains B and C. Activity is again initiated at the beginning of chain A by external current injection. Thus each run of this simulation gives rise to either the sequence AB or AC. The transition probability is defined by the fraction of simulations resulting in AB divided by the total number of simulations. We examine three cases in regards to auditory feedback. In the first case, there is no auditory feedback. In the second case, auditory feedback from syllable A targets chain B, biasing the network towards producing AB. In the final case, auditory feedback from syllable A targets chain C, biasing the network away from AB and toward AC.

In some cases (especially at the lowest temperatures), occasionally a simulation would result in the synfire chain signal dying out (i.e. failing to propagate from one group to the next), most often at a branch point. These simulations are disregarded for computing transition/repeat probabilities. In almost all scenarios, this outcome occurred less than 2% of the time. The only exception was the branch point network with no auditory feedback at the lowest temperature ($\Delta T = -4°C$), where it occurred on ~7% of the simulations due to the combination of a large temperature change and no auditory feedback to provide extra excitatory drive to the chains.

## ACKNOWLEDGEMENTS

The authors wish to acknowledge Guerau Cabrera and Bruce Langford for their assistance at the initial stage of the project and Dmitry Aronov for technical advice. This work was supported by NSF Grant 0827731, the Pennsylvania State University Department of Physics and the Huck Institute for Life Sciences.

FIGURE LEGENDS

**Figure 1**. Temporal structures of the Bengalese finch song and the effects of HVC temperature on their durations.
(a) A song bout consists of a sequence of song syllables (labeled with letters 'a' through 'k' above the spectrogram).
(b) The syllables are separated by silent gaps.
(c) The song syntax can be visualized by a transition diagram. The arrows show the allowed transitions between the syllables, and the numbers are the transition probabilities.
(d) Decreasing HVC temperature increases the syllable and the gap durations, while increasing the temperature reduces the durations. As an example, a song segment consisting of four syllables are shown at different HVC temperatures, aligned to the onset of the first syllable.
(e) Histograms of the temperature-induced fractional duration changes of all syllables (left) and gaps (right) in our dataset. Syllable durations are stretched by -2.8±0.9 %/°C (s.d., N=34) and gap durations are stretched by -4.2±2.5 %°C (s.d., N=50).

**Figure 2**. Effects of HVC temperature on the syllable repeats.
(a) Decreasing HVC temperature increases the length of the repeat of syllables. An example is shown. Following syllable 'C', syllable 'A' repeats variable number of times and transitions to 'B'. Representative repeat segments at different HVC temperatures are shown. The segments are aligned to the onset of syllable 'C'.
(b) Histograms of the repeat lengths of syllable 'A' at different HVC temperatures. The temperature change is indicated on the top of each plot. Cooling systematically shifts the distribution, decreasing the mean number of repeats.
(c) Repeat length of Type I (long and variable) and Type II (short and stereotyped – see text for precise definition) repeated syllables at different temperatures, centered by the mean across all temperature conditions. Dash lines are the linear fit for these two types. Temperature has more effects on the Type I syllables than the Type II syllables ($p=6.6\times10^{-8}$, multivariate regression analysis).
(d) Repeat length of all Type I syllables from 5 birds at different temperatures (N=8). Error bars are s.e.m.. Colors indicate different individuals.
(e) Compared to the repeat length distribution derived from the time-conservation model at $\Delta T = -2°C$, the observed effect of cooling HVC by 2°C shows greater reduction of repeat length ($p=1.3\times10^{-11}$, t-test, $N_{model} = 1158$, $N_{observation} = 179$).
(f) Repeat length of the Type I syllables and the corresponding simulated length based on the time-conservation model. The model-generated effect of HVC temperature on repeat length is significantly weaker than the real data ($p=2.1\times10^{-5}$, multivariate regression analysis).

**Figure 3** Effects of HVC temperature on the branch points in the song syntax.
(a) An example of a branch point affected by HVC temperature. Top: spectrograms of song segments showing that syllable 'K' can be followed by either syllable 'B' or syllable 'D'. Bottom: the transition probabilities from 'K' to 'B' or 'D' under control and 4°C cooling conditions.
(b) Dependence of the transition entropy on HVC temperature for all branch points. Entropies are normalized by their maximum possible values. Red curves are branch points that significantly decrease in transition entropy as temperature increase; blue curves significantly increase transition entropy with temperature; and grey curves show branch points that does not significantly change with temperature (t-test on slope). Transition entropies at $\Delta T = 0$ are within the box and are used to obtain plot (c).
(c) Slopes of the normalized entropy vs $\Delta T$ curve plotted with respect to the normalized entropy at $\Delta T = 0$. The slope is significantly greater than 0 ($p=1.2\times10^{-5}$, t-test).

**Figure 4** Effects of RA temperature on song tempo and syntax.
(a) Mean fractional stretch of syllables (N=22) and gaps (N=34) at various RA temperatures compared to the stretch with changing the temperature of HVC. Error bars are s.e.m..
(b) Repeat length of the four Type I repeated syllables at different RA temperatures. Error bars are s.e.m..
(c) Comparison of the temperature effects on repeat length with cooling HVC and RA. Changes of RA temperature yield much less variation of repeat length ($p=2.2\times10^{-7}$, multivariate regression analysis).
(d) Transition entropy of the seven branch points from two birds at different RA temperatures. All curves have slopes that are not significantly different from zero.
(e) Slope of the curves with respect to the normalized entropy at $\Delta T = 0$. The slopes are independent of the transition entropy at $\Delta T = 0$ (p=0.21, t-test).

**Figure 5** Modeling the effects of HVC cooling on song syntax
(a) In the model, the temporal structure of the syllables is encoded in chain networks in HVC – one for each syllable. Feedforward connections between groups of RA-projecting neurons give rise to stereotyped propagation of excitation through the chain which drives production of the syllable through group-specific projections to motor neurons in RA. In the diagram, the red neurons/connections show how activity in one group of neurons in HVC simultaneously drives motor neuron activity downstream in RA and activates the next group of HVC neurons.
(b) Syllable durations are longer at cooler temperatures. Dots are averages and bars are standard deviations across multiple trials (number of trials: N = 2000 at $\Delta T$ = -4°C; 1,000 at -2°C; 500 at 0°C and 2°C). The trend is approximately linear (red line is best linear fit).
(c) To examine the effects of temperature on syntax, the model is extended to produce probabilistic transitions between syllables. Branching connections between chains restrict the allowable syllable transitions and their relative strengths determine transition probabilities. Syllable-specific auditory signals can target specific chains, allowing auditory feedback to play an online role in shaping transition probabilities. Stimulus-specific adaptation of these auditory signals is modeled, as it has been shown to be sufficient to reproduce salient features of the syntax. Here we show a scenario for producing a Type II repeat syllable (long and variable repetitions), 'A'. Auditory feedback from 'A' targets chain-A, driving long repetition bouts. Stimulus-specific adaptation eventually quenches this positive feedback loop, causing termination of the repetition.
(d) The number of repetitions for such a syllable as a function of temperature. Dots are averages and bars are standard deviations across multiple trials (number of trials the same as in Fig. 5b). The trend is approximately linear (red line is best linear fit). The dashed purple line shows the expected change in repetition from an alternative time-conservation model where HVC does not play an active role in syntax encoding and the decrease in repetition is merely a side-effect of the longer duration syllables. Our model predicts a much larger change in repetition compared to this alternative explanation.

(e) Repeat distributions for this syllable. Blue bars are the repeat distribution at each temperature, while the gray bars are the distribution at $\Delta T = 0°C$ for easy comparison (compare to Fig. 2b).
(f) A scenario for examining the impact of temperature on branch points. Here chain-A branches to chain-B and chain-C. If no auditory feedback is present, A-to-B and A-to-C connection strengths are chosen so that 'AB' and 'AC' are approximately equally likely outcomes of simulating this network. However auditory feedback from 'A' can bias this transition by targeting either chain-B or chain-C, thus making 'AB' or 'AC' (respectively) more likely. The diagram illustrates the latter case.
(g) The entropy of the branch point in three cases: (1) no auditory feedback present (black line); (2) auditory feedback provides bias toward 'AB' (blue curve); (3) auditory feedback provides bias toward 'AC' (purple curve). Dots are entropies and bars are 90% confidence intervals (computed first on transition probabilities using Wilson's confidence interval for proportions with a continuity correction and then propagated to the entropy). The dashed red line is the maximum possible entropy for such a binary random variable (1 bit). Bias provided by auditory feedback causes a decrease in entropy at the normal temperature. Cooling causes the syntax to revert to a more random (higher entropy) state.

**Figure 6** DAF effect on syllable repeats.
(a) Top: a segment of intact song with a long repeat. Bottom: Disturbed song from the same bird.
(b) Repeat length distribution without (top) and with (bottom) DAF. DAF significantly reduces repeat length (p=5.1×10$^{-13}$, t-test, $N_{control} = 104$, $N_{DAF} = 106$, one type of repeated syllable from Bird 7).
(c) The p-values of the KS-test between repeat length with DAF and with different temperature changes. At $\Delta T = -2°C$, the similarity between cooling and DAF is largest, p=0.07. Zoomed graph shows the comparison of the repeat length distribution with DAF (N=106) and with cooling HVC by -2°C (N=54).

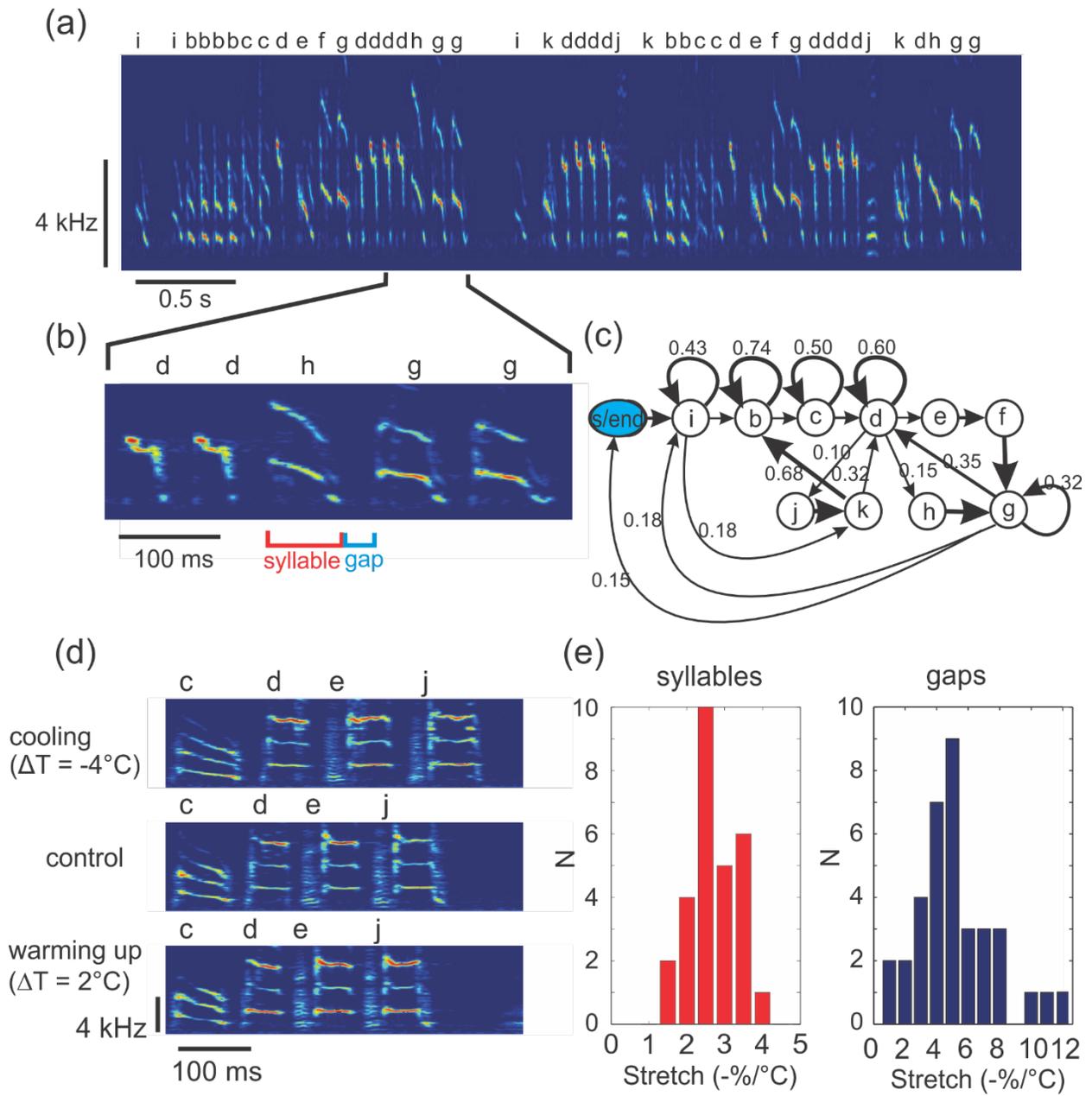

Figure 1     Zhang et al (2015)

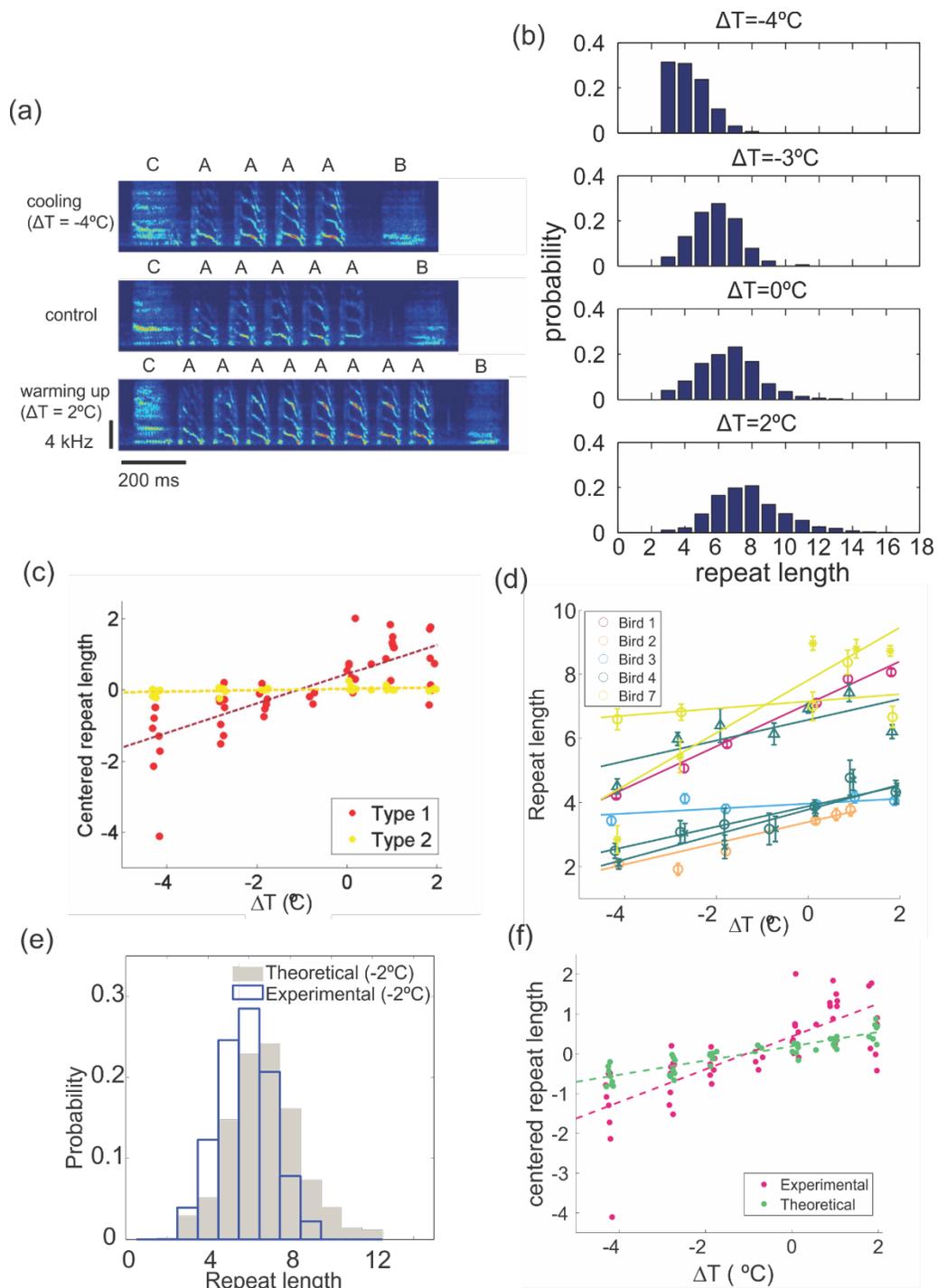

Figure 2 Zhang et al (2015)

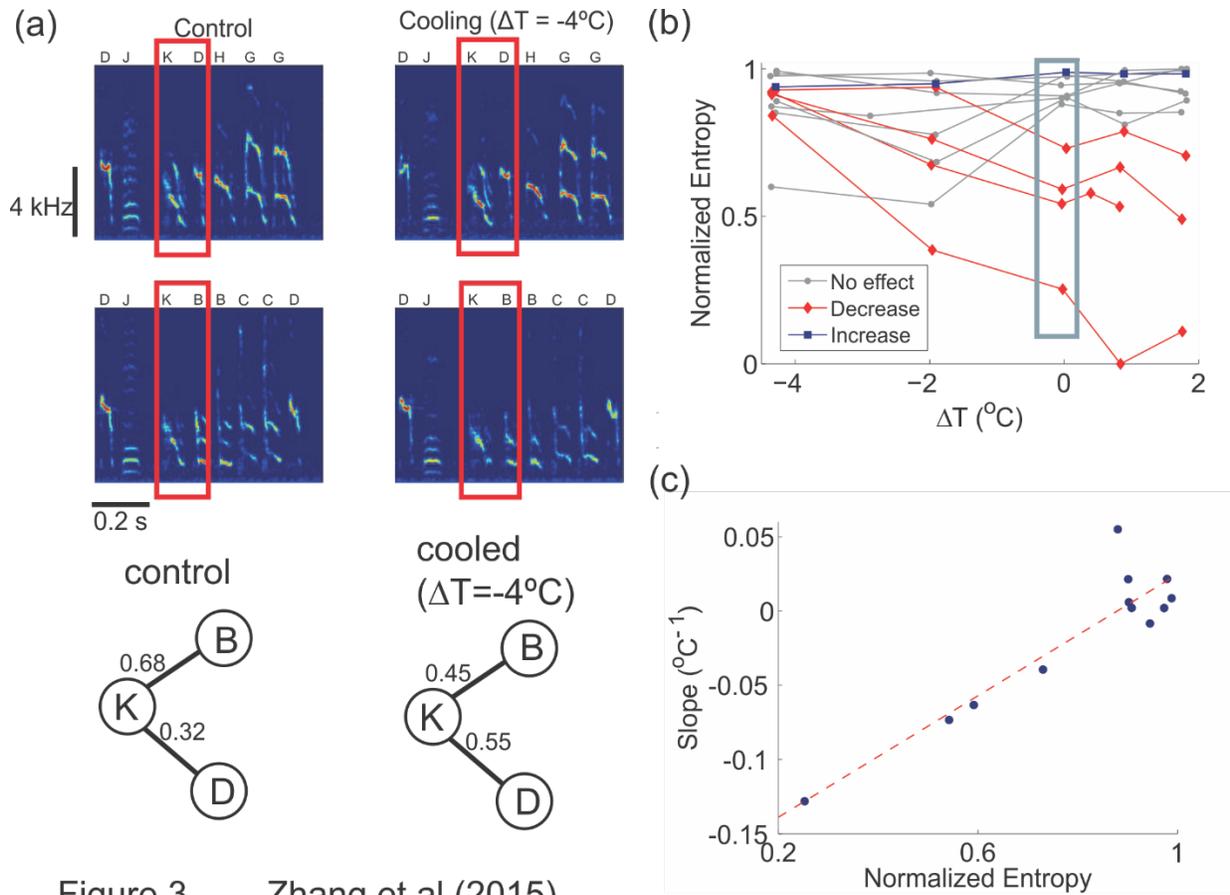

Figure 3    Zhang et al (2015)

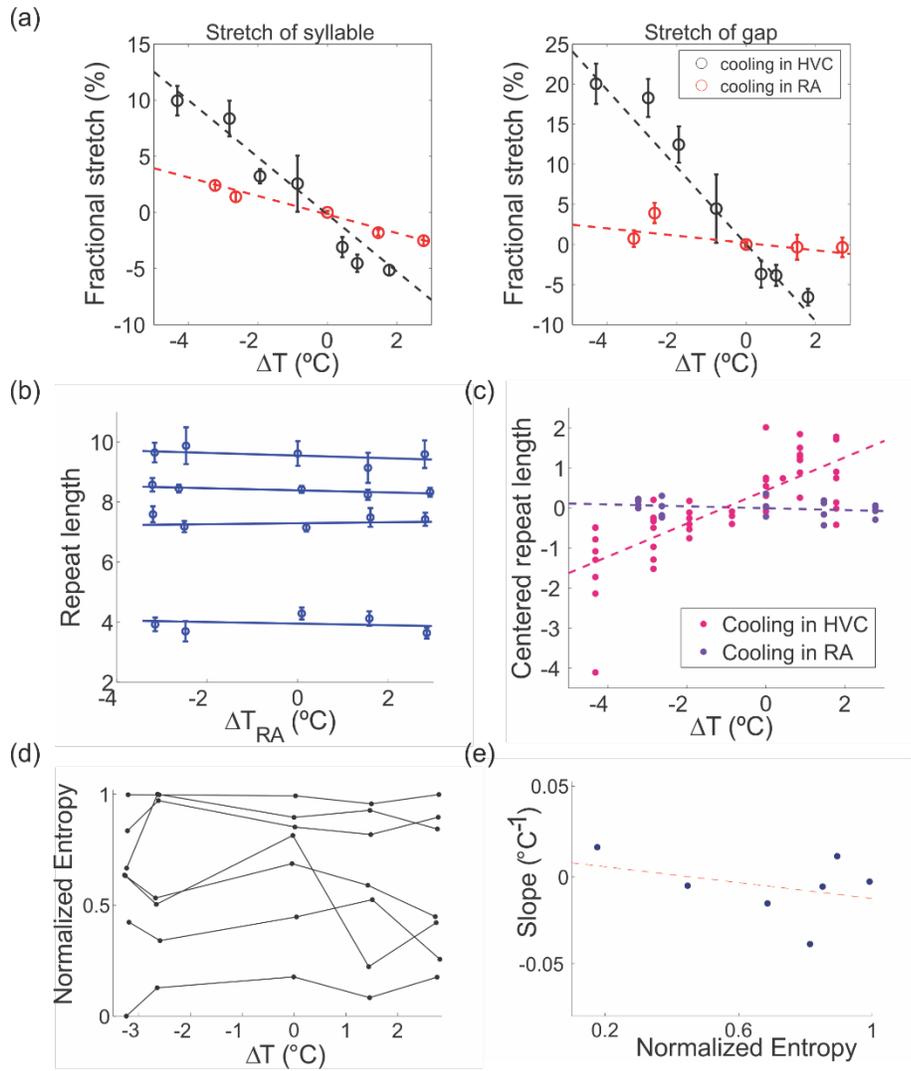

Figure 4    Zhang et al (2015)

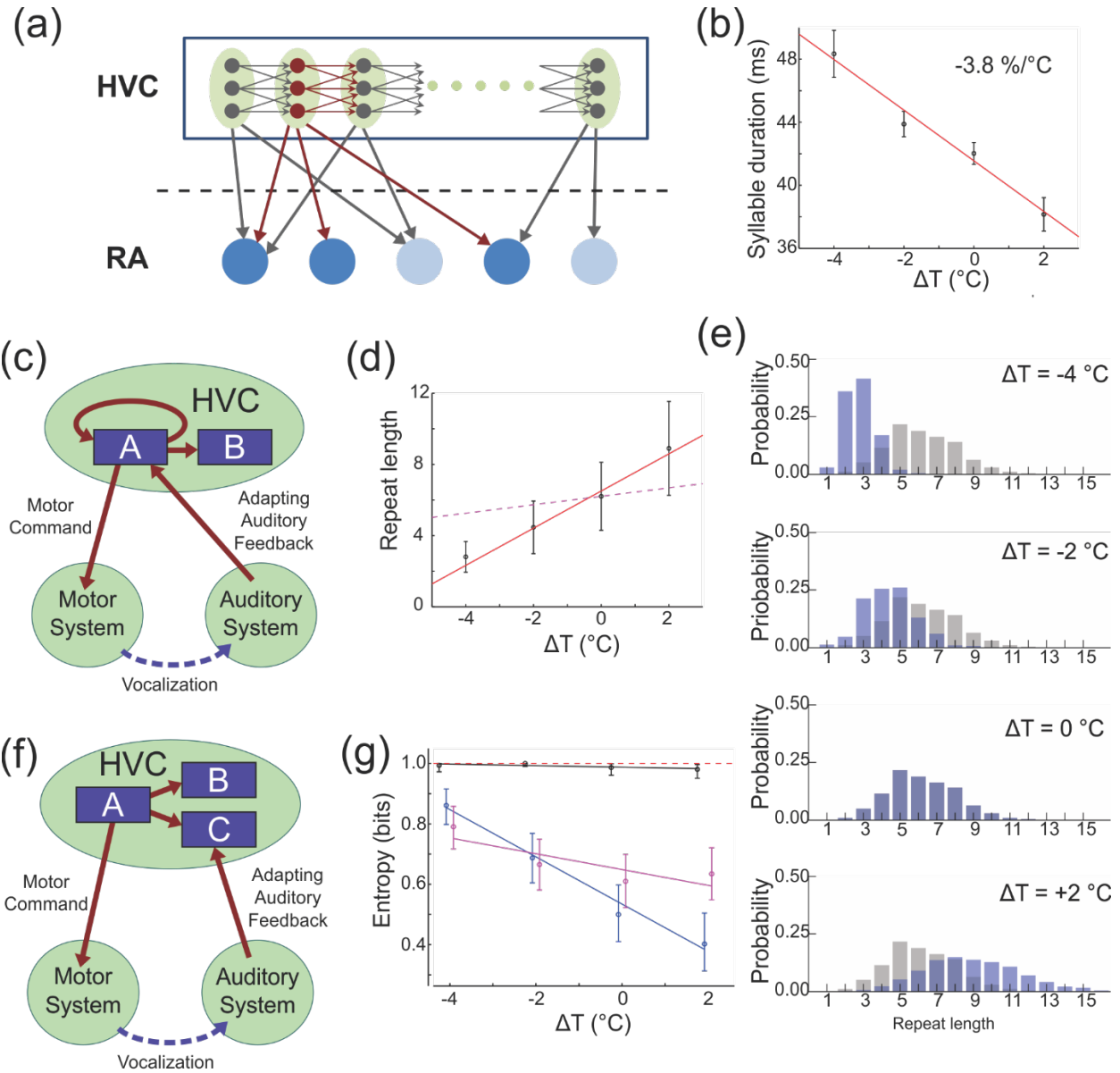

Figure 5 Zhang et al (2015)

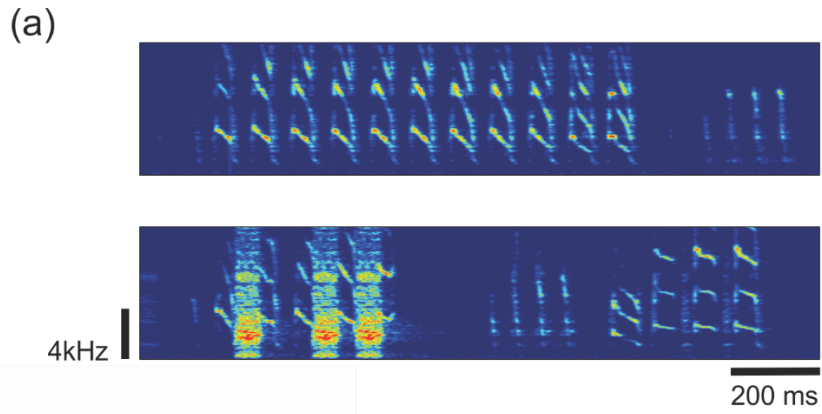

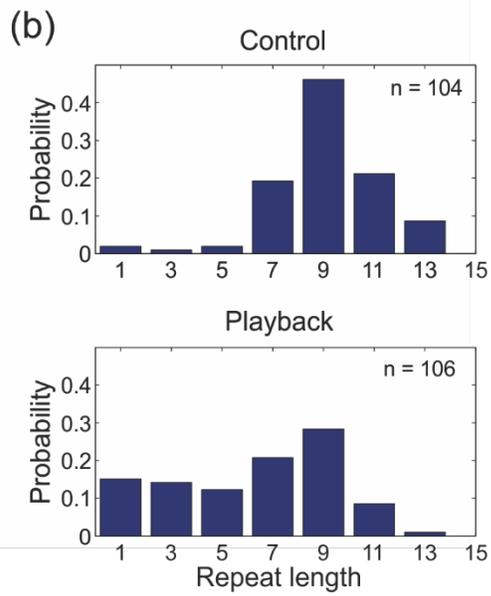

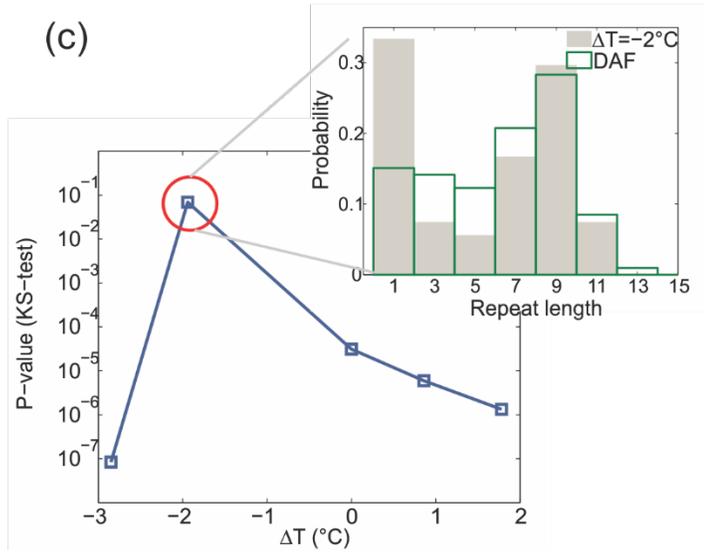

Figure 6    Zhang et al (2015)